\documentclass{JHEP3}

\usepackage{epsfig,bbm,bm,amssymb,graphicx}

\author{Gonzalo A. Palma \\Instituut-Lorentz for Theoretical Physics, \\Universiteit Leiden, \\P.O. Box 9506, NL-2300 RA Leiden,\\The Netherlands\\ \\email: palma@lorentz.leidenuniv.nl}

\author{Subodh P. Patil \\Humboldt Universit\"at zu Berlin,\\ Institut f\"ur Physik,\\Newtonstra{\ss}e 15, D-12489 Berlin, \\Germany\\ \\email: subodh@physik.hu-berlin.de}

\date{\today}
\preprint{HU-EP-08/45}

\title{Inflation with a stringy minimal length, reworked}

\newcommand{\eq}[2]{\begin{equation}\label{#1}{#2}\end{equation}}
\def\bea{\begin{eqnarray}}
\def\eea{\end{eqnarray}}
\def\be{\begin{equation}}
\def\ee{\end{equation}}
\def\ba{\begin{array}}
\def\ea{\end{array}}

\abstract{In this paper we revisit the formulation of scalar field theories on de Sitter backgrounds subject to the generalized uncertainty principle (GUP). The GUP arises in several contexts in string theory, but is most readily thought of as resulting from using strings as effective probes of geometry, which suggests an uncertainty relation incorporating the string scale $l_s$. After reviewing the string theoretic case for the GUP, which implies a minimum length scale $l_s$, we follow in the footsteps of Kempf and concern ourselves with how one might write down field theories which respect the GUP. We uncover a new representation of the GUP, which unlike previous studies, readily permits exact analytical solutions for the mode functions of a scalar field on de Sitter backgrounds. We find that scalar fields cannot be quantized on inflationary backgrounds with a Hubble radius $H^{-1}$ smaller than the string scale, implying a sensibly stringy (as opposed to Planckian) cutoff 
 on the
  scale of inflation resulting from the GUP. We also compute $(H l_s)^2$ corrections to the two point correlation function analytically and comment on the future prospects of observing such corrections in the fortunate circumstance our universe is described by a very weakly coupled string theory.}

\begin{document}

\section{Introduction}

It is perhaps common lore at present that whatever the modifications to geometry that arise in any consistent formulation of quantum gravity, the effective degrees of freedom of geometry and/or matter will organize themselves such that a minimal length scale becomes manifest at high enough energies. Indeed, some phenomenological approaches take a minimum length scale as a prior and work from the bottom up with models which have such a feature built in. One particular example is non-commutative field theory (see~\cite{ncft} for a review), which explores the consequences of field theories defined on spacetimes where the coordinates satisfy non-trivial structure relations, for example
\eq{ncft}{[x^\mu,x^\nu] = i\theta^{\mu\nu}I,}
which, long before having made an appearance as the structure relations satisfied by the endpoints of open strings in non-trivial RR flux backgrounds~\cite{flux1, flux2}, appeared in the pioneering work of Snyder~\cite{snyder}. A direct corollary of~(\ref{ncft}) is the uncertainty between any two spatial coordinates $x,y$ for which the component $\theta^{xy}$ is non-vanishing
\eq{sunc}{\Delta x\Delta y \geq |\theta^{xy}|,}
which implies a minimum amount of localization one can simultaneously achieve for all spatial coordinates. This calls into question the very notion of a spatial point and implies an effective minimum resolution 
\eq{min}{\Delta x \sim \sqrt{|\theta|},}
where $x$ is any given spatial coordinate and $\theta$ is a typical entry of the non-commutativity tensor. 

In string theory the last example arises dynamically in certain specific contexts (mentioned above) rather than as a prior input, and is to be viewed not as a fundamental property of spacetime, but rather as one of many examples of modifications to classical notions of geometry that arise dynamically when we probe spacetime at high energies with strings.\footnote{A more dramatic example is offered by various matrix models, which suggest that spacetime itself is generated through the dynamics of strings at high energies (see~\cite{seiberg} for an overview).} Yet even though the nature of the modifications to classical geometry implied by string theory typically depends on the dynamical regime we find ourselves in, from an effective field theory point of view, there is one modification that appears to be generic, encoded by the generalized uncertainty principle (GUP)
\eq{gup}{\Delta x \Delta p \geq \frac{1}{2}[1 + \beta(\Delta p)^2 ],}
where $\beta$ is a parameter of dimension length squared, which will turn out to be commensurate to $\alpha'$. The GUP can be thought of as arising from the studies of string scattering at very high energies~\cite{mende, ven} or from worldsheet renormalization group considerations~\cite{konishi}. There is, however a quick and heuristic way to understand the GUP, which first requires us to appreciate the fact that~(\ref{gup}) implies a minimum spatial resolution. By solving for $\Delta p$ once we saturate the bound in the above, we find that
\eq{dp}{\Delta p = \frac{2}{\beta}[\Delta x \pm \sqrt{(\Delta x)^2 - \beta}],}
which, after noting that $\Delta p$ is an intrinsically real and positive quantity by definition, implies
\eq{mx}{\Delta x \geq \sqrt \beta \sim \sqrt{\alpha'} \sim l_s.}
We can thus understand the GUP as the result of probing spacetime events by scattering strings off of them. To increase the spatial resolution, one must fire in strings with higher and higher center of mass momentum $P_{\rm c.o.m.}$. However, the mass shell Virasoro constraint~\cite{pol} implies that for large enough energies\footnote{Where we assume in the case of closed string probes, any left moving excitations $\tilde N$, if any have been solved in terms of $N$ using the level matching constraint.}
\eq{vc}{P_{\rm c.o.m.}^2 \sim N,}
where $N$ is the number of excited string modes. Hence, we see that as we dial up the center of mass momentum of our string probes, we will also be exciting more and more oscillator states. Since more oscillators imply more wiggles to our string probe, we see that we cannot go on increasing the center of mass momentum indefinitely without at some point encountering the competing effect causing us to lose spatial resolution through exciting these oscillators. At around this string scale, we expect this effect to be very strong, and it is precisely this behavior that is captured by the GUP, which renders the string length to be the effective minimal length if strings are our only probes of geometry.\footnote{One can imagine attempting to evade this limitation by probing spacetime with lower dimensional objects, i.e. D0-branes. However, as discussed above, these probes also exhibit the string length as a critical scale in certain backgrounds below which the nature of spacetime i
 s distinctly non-classical~\cite{seiberg} (see also~\cite{sethi, london, erd} for using these dynamics to probe cosmological singularities and possibly resolve them).}

The goal of this paper is to explore the consequences of the GUP for a scalar field theory defined in particular, on a de Sitter background. Our motivation is inflationary physics, and the possibility of this minimal length having imprinted itself on the cosmic microwave background, as has been previously considered by several authors~\cite{kempf1, Kempf:2001fa, guptp, kempf2}. Just as previous work on this subject, our report is in the spirit of an investigation of a toy model which incorporates a stringy minimal length, and therefore we take (\ref{gup}) 
to be exact even though in reality they represent only the first terms in a series expansion in $\alpha'$ (and is a toy model for this reason).\footnote{Other consequences of taking (\ref{gup}) literally for laboratory scale quantum phenomena were explored in \cite{vagenas}. An interesting study that employs similarly modified commutation relations and explores their consequences for reproducing features of gravity was studied in \cite{Jackson:2005ue}.}

After retracing the footsteps of Kempf~\cite{kempf1} and concerning ourselves with self consistently formulating the GUP (\ref{gup}) and looking for suitable operator representations, we uncover a new representation of the relations~(\ref{gup}) which permits a new class of exact, analytical solutions for the mode functions of a massless scalar field theory on de Sitter space. These mode functions relate to the usual mode functions of a massless scalar field (used to define the Bunch-Davies vacuum) in an obvious way, and exhibit various interesting features. In particular, we uncover the (only to be expected) result that one cannot write down normalizable mode functions if the scale of inflation $H$ is such that $\beta H^2 > 1$. This simply encapsulates the fact that effective field theory is breaking down as one approaches spacetime curvatures comparable to the string scale, implying a cutoff on the scale of inflation coming from the GUP, which at weak string coupling, is sub
 -Planckian. We then proceed to derive $H^2/m^2_s$ corrections to the two point correlation functions and discuss corrections to various other observables of interest coming from the CMB. We comment on the (charitable) circumstances under which these modifications may be observable, and point out how in the presence of a cutoff, bounds coming from particle production at the end of inflation previously used to rule out non trivial alpha vacua, are relaxed somewhat (although stability arguments used to rule them out still apply). The significance of our results lie in the fact that unlike previous investigations, we have obtained exact solutions to the mode equations of a scalar field theory in de Sitter space subject to the GUP with minimal fuss, and in addition to yielding several new results, will hopefully facilitate future investigations of the effects of string physics on the CMB via the GUP.

The outline of this report is as follows: We begin by formulating the generalized uncertainty principle on Minkowski and de Sitter spacetimes, reviewing the auxiliary representation first introduced by Kempf~\cite{kempf1} and discussing the modified mode equations for a scalar field in such a representation. We then present a new representation of the GUP, where the treatment of quantizing a scalar field of such a background proceeds much more simply and yields exact solutions to the mode equations that permits an exact calculation of the two point correlation of a free scalar field. We discuss how the approach we present here avoids some of the operator ordering ambiguities associated with previous approaches. We then discuss various consequences for inflationary and cosmological physics, and the possibility that such effects (although they appear as $H^2/m_s^2$ corrections) might appear at weak string coupling, after which we comment on how our results fits in to the wider 
 literature on the `trans-Planckian problem' and offer our concluding thoughts.

\section{Generalized uncertainty principle on Minkowski backgrounds}

We commence by reviewing the treatment given by Kempf in~\cite{kempf1} for the case of Minkowski backgrounds. Consider the situation in which the standard commutation relations $[X^i , P_j] = i \delta^i_j$ are modified for large values of the momenta, in the following manner:
\be
[X^i,P_j]= i \left( f(P^2) \delta^i_j + g(P^2) P^i P_j \right). \label{mod1}
\ee
Here $f$ and $g$ are functions of $s=P^2$ to be determined. Self consistency, via the Jacobi identities then imposes the following relation between $g$ and $f$
\be
g(s) = \frac{2 f  f'}{f - 2 s f'}, \label{relation-f-g}
\ee
where $'= \partial_{s}$. A simple choice for these functions which satisfies Eq.~(\ref{relation-f-g}) consists of $g = 2 \beta$ and $f(s) = 2 \beta s /(\sqrt{1+4 \beta s} - 1)$ \cite{kempf1}. It is easy to check that this choice reproduces~(\ref{mod1}), and reduces to the usual canonical commutation relations for small momenta. One may then construct a field representation of the new commutation relation~(\ref{mod1}) with this choice for $f$ and $g$ by introducing as in~\cite{kempf1}, a set of auxiliary variables $\rho_i$, such that
\be
X^i \phi(\vec \rho)  = i \frac{\partial}{\partial \rho_i} \phi(\vec \rho),  \qquad  {\rm and } \qquad
P_i \phi(\vec \rho)  = \frac{\rho_i}{1-\beta \rho^2} \phi(\vec \rho) , \label{X-P-rho} 
\ee
which are symmetric though not self adjoint (see~\cite{kempf1} for a discussion of this and related points concerning the mathematical sense of this representation) with respect to the scalar product:
\be
(\phi_1, \phi_2) = \int_{\beta \rho^2<1}  \!\!\! d^3 \rho \, \phi_1^{*}(\vec \rho) \phi_2(\vec \rho).  \label{scalar-prod}
\ee
We see that it is necessary to impose the boundary condition $\phi(\rho=\beta^{-1/2}) = 0$, where $\rho \equiv |\vec \rho|$, in order for this representation to be well defined. The requirement of having a real field imposes the additional reality condition $\phi^{*}(t,\vec \rho) = \phi(t,- \vec \rho)$. We parenthetically note that the $P_i$'s are no longer generators of spatial translations. Instead one may think of the new variables $\rho_{i}$ as the generators of translation, in the sense that:
\be
[X^i, \rho_j] = i \delta^i_j.
\ee

Beginning  with Minkowski space, the basic ansatz proposed in ref.~\cite{kempf0} and elaborated upon in ~\cite{kempf1} towards constructing scalar field theories incorporating the effects of~(\ref{mod1}) is to retain the validity of the following paraphrasal of the usual free field action
\be
\label{presc}
S = -\frac{1}{2} \int dt  \left[  (\phi, \partial_{t}^2 \phi)  +  (\phi, {\bf P}^2 \phi)  \right],
\ee
where the scalar product is as in~(\ref{scalar-prod}).
In the $\rho$-representation, we then see that the representation~(\ref{X-P-rho}) results in the modified action
\be
S = \frac{1}{2} \int_{\rho^2 \beta < 1} dt \, d^3 \rho \left[  | \dot \phi |^2 - \frac{\rho^2}{(1 - \beta \rho^2)^2} |\phi |^2 \right], \label{action-Mink}
\ee
and yields the following equation for the mode functions:
\eq{modfe}{\ddot\phi(t,\rho) + \frac{\rho^2}{(1 - \beta\rho^2)^2}\phi(t,\rho) = 0.}
This is easily solved in terms of positive and negative frequencies modes $\phi(t,\rho) = e^{\pm i \omega( \rho ) t}$ but now with the modified dispersion relation of the form:\footnote{Although the group velocity $v = \partial \omega / \partial \rho$ deduced from this dispersion relation features superluminal states $v>1$, here causality is not violated as the construction of closed non-causal paths in spacetime would require an infinite amount of energy.}
\be
\omega(\rho) = \frac{\rho}{1 - \beta \rho^2}. \label{mod2}
\ee
Notice that in position space, one may express these solutions in terms of the usual plane waves
$\phi(t,x) = A e^{i\omega(\rho) t - i\rho\cdot x} + B e^{-i\omega(\rho) t + i\rho\cdot x}$.
In order to quantize these solutions one has to impose the canonical commutation relation:
\be
[\phi(\tau,\rho), \pi(\tau,\rho')] = i \delta^{3}(\rho - \rho'). \label{canonical-com}
\ee
where $\pi(t, \rho) = \dot \phi (t, -\rho)$ is the canonical momentum deduced from the action (\ref{action-Mink}). This can be done by expressing $\phi$ in terms of creation and annihilation operators $a^{\dag}(\vec \rho)$ and $a(\vec \rho)$ satisfying the standard commutation relation $[a(\rho),a^{\dag}(\rho')] = \delta^{3}(\rho-\rho')$.\footnote{Although the phase space coordinates satisfy the GUP, field quantization is to proceed in the usual canonical manner. The point is that the GUP is not a modification of quantum mechanics per se, but rather captures the effects of using strings as probes of geometry. This can also be seen from the fact that all derivations of the GUP (c.f. \cite{mende, ven, konishi}) rely on first quantizing strings, where quantization on the worldsheet proceeds in the usual canonical manner.} The result is
\be
\phi(t, x) = \int_{\rho^2\beta < 1} d^3\rho \Bigl[ \frac{e^{i \omega t - i\rho\cdot x}}{\sqrt{2 \omega}} a^{\dag}(\vec \rho)  +  \frac{e^{-i \omega t + i\rho\cdot x}}{\sqrt{2 \omega}} a(-\vec \rho)\Bigr],
\ee
where the factor $1/\sqrt{2 \omega}$ is the familiar normalization factor appearing in Minkowski backgrounds, appearing after imposing Eq.~(\ref{canonical-com}).

\section{Generalized uncertainty principle on FRW backgrounds}

The main purpose of this section is to extend the previous results to FRW-backgrounds by proposing a new representation of the generalized uncertainty principle on de Sitter backgrounds. To make our approach clear and distinguish it from previous results, we first review the ansatz proposed in ref.~\cite{kempf1} and comment on some difficulties posed by such a scheme. 

\subsection{Physical coordinate ansatz}

To start with, recall that spatially flat FRW-backgrounds may be described by the following metric
\be
ds^2 = a^2 (\tau) \left( - d \tau^2 +  \delta_{i j} d x^i d x^j   \right), \label{FRW-metric}
\ee
where $\tau$ is the usual conformal time and $x^i$ denotes comoving coordinates. In the absence of GUP the action for a free scalar field is given by:
\be
S = - \frac 12 \int d\tau d^3 x \, a^2 (\tau) \left[  (\partial_{\tau} \phi)^2 - \sum_{i=1}^{3} (\partial_{x^i} \phi )^2  \right]. \label{FRW-com-coord}
\ee
Since the generalized uncertainty principle of eq.~(\ref{mod1}) is spelled in terms of proper distances, it is then useful to consider the previous action written in terms of physical spatial coordinates  $y^i = a x^i$ rather than in comoving coordinates. One finds
\be
S = - \frac 12 \int d\tau d^3 y \, a^{-2} (\tau) \left[  \left( A \phi  \right)^2 - a^{2}(\tau) \sum_{i=1}^{3} (\partial_{y^i} \phi )^2  \right], \label{FRW-phys-coord}
\ee
where
\be
A =  \partial_{\tau} + i \frac{a'}{a}  P_i y^i - 3 \frac{a'}{a} , \qquad \mathrm{and} \qquad P_i = - i \partial_{y^i} . \label{operator-A}
\ee
Notice that $A$ can be thought of as a convective derivative taking into account the fact that we are now in a non-comoving frame. In terms of the internal product, the previous action can in fact be written as
\be
S = - \frac{1}{2} \int \frac{d \tau}{a} \left[ ( \phi, A^{\dag} A \phi ) + a^2 \sum_{i=1}^3 ( \phi , {\bf P}^2 \phi )  \right]. \label{FRW-action-1}
\ee
The ansatz of ref.~\cite{kempf1} consisted of promoting at this stage all of the operators $P_i$ and $Y^i$ of Eqs.~(\ref{operator-A}) and~(\ref{FRW-action-1}) to satisfy the generalized commutation relations~(\ref{mod1}), with $X^i$ replaced by $Y^i$, and again introduce the auxiliary variable $\rho_i$ as in~(\ref{X-P-rho}). The resulting action was found to be:
\be
S =  \frac{1}{2} \int_{\beta \rho^2<1} d \tau  \, d^3 \rho \, \frac{1}{a} \left[ \left|  \left( \partial_{\tau} -  \frac{a'}{a}  \frac{\rho_i}{1 - \beta \rho^2} \partial_{ \rho_i} - 3 \frac{a'}{a} \right) \phi \right|^2   - \frac{a^2 \rho^2 |\phi |^2}{(1 - \beta \rho^2)}   \right]. \label{FRW-action-2}
\ee
At this point one might worry about the appearance of $y^i$ in Eq.~(\ref{FRW-action-1}) via~(\ref{operator-A})  [or equivalently $\partial_{ \rho}$ in  (\ref{FRW-action-2})] which causes translational invariance of the system to be no longer manifest. In addition, the presence of the $\partial_{ \rho}$ terms in the above couples different $\rho$-modes. By making the change of variable 
\eq{cov}{k^i = a\rho^i e^{-\beta\rho^2/2},}
one can actually decouple these modes and derive the following equation of motion for the transformed $k$-modes \cite{kempf1}
\eq{kmod}{\phi''_k + \frac{\nu'}{\nu}\phi'_k + \Bigl[\mu - 3\Bigl(\frac{a'}{a}\Bigr)' - 9\Bigl(\frac{a'}{a}\Bigr)^2 - \frac{3a'\nu'}{a\nu} \Bigr]\phi_k = 0,}
with
\begin{eqnarray}
\label{mu}
\mu(\tau,k) &:=& - \frac{a^2 W(-\beta k^2/a^2)}{\beta(1 + W(-\beta k^2/a^2))^2} = \frac{a^2\rho^2}{(1 - \beta\rho^2)^2},\\
\label{nu}
\nu(\tau,k) &:=& \frac{e^{-\frac{3}{2}W(-\beta k^2/a^2)}}{a^4(1 + W(-\beta k^2/a^2))} = \frac{e^{\frac{3}{2}\beta\rho^2}}{a^4(1 - \beta\rho^2)},
\end{eqnarray} 
where $W(x)$ corresponds to the Lambert W-function, defined as the inverse of the function $x e^x$. The solutions to these mode equations were solved numerically and semi-analytically in~\cite{guptp, kempf2}. As is clear from~(\ref{kmod}), these solutions are rather involved and we do not discuss them further, as we wish instead to present a new class of solutions.

\subsection{An alternative prescription}

We note that when $\beta = 0$, there are no ordering ambiguities in writing the operator $A$ as we do in eq.~(\ref{operator-A}). To see this, observe that instead of eq.~(\ref{operator-A}) we could very well have introduced an arbitrary real function $\sigma = \sigma(y,P)$, and write more generally:
\be
\label{ordam}
A =  \partial_{\tau} + \frac{a'}{a}  \left[ i  P_i y^i - 3 \right] \sigma(y,P) + \left[ 1 - \sigma(y,P) \right] \frac{a'}{a} i   y^i P_i  .  \label{operator-A2}
\ee
Since $[Y^i , P_j] = i \delta^i_j$, the presence of $\sigma(y,P)$ plays no physical role whatsoever. However, as first observed in ref.~\cite{Ashoorioon:2004vm}, this is no longer the case if we allow $\beta$ to be non-zero and promote $P_i$ and $y^i$ to satisfy the new commutation relations. In this case, the function $\sigma(y,P)$ cannot be eliminated from the action, manifesting the ordering ambiguity in this approach. It was further noticed in ref.~\cite{Ashoorioon:2004vm} that this ordering ambiguity also  breaks the equivalence between previously equivalent choices of gauge for the scalar modes of a scalar field theory coupled to gravity. Specifically, when we consider the quadratic action for the curvature perturbations $\mathcal R$, we have to change variables to the Mukhanov variable $u = -z\mathcal R$ in order to cast the action in canonical form.\footnote{In conformal time for example, $z = a^2\phi'_0/a'$, where $\phi_0$ is the background scalar field.} When $\beta
  = 0$, the two actions expressed in different gauges differ by a boundary term. When $\beta \neq 0$, this difference no longer corresponds to a boundary term, and this ambiguity is related to a particular choice of the function $\sigma(y,P)$.

The physical interpretation of the ordering ambiguity inherent in (\ref{ordam}) is straightforward. Observe that when $\beta = 0$, there is no distinction between the momentum operators $P_i$ and generators of spatial translations $\rho_i = - i \partial_{y^i}$. At finite $\beta$ this degeneracy is broken and we are forced to distinguish between them. It should be clear that at $\beta = 0$, the only reason why we are allowed to write down $A$ as in eq.~(\ref{operator-A}) is because $P_i$ is strictly acting as a generator of translation (after all, the convective derivative $A$ (\ref{operator-A}) in eq.~(\ref{FRW-phys-coord}) appears as a result of
  a rewriting the action in a different coordinate frame). However at finite $\beta$, as we have argued above, an ordering ambiguity appears. In order to avoid the ambiguity at finite $\beta$, we might consider instead $\rho_i$ in place of $P_i$:
\be
A =  \partial_{\tau} + i \frac{a'}{a}  \rho_i y^i - 3 \frac{a'}{a} , \qquad \mathrm{and} \qquad \rho_i = - i \partial_{y^i} . \label{operator-A-2}
\ee
Notice that $A$ is now independent of $\beta$, and therefore, this alternative prescription accomplishes two important things: First, it eliminates the presence of the unphysical function $\sigma(y,P)$ which appears as a consequence of writing down the theory in a non-coordinate basis. And second, it ensues that the action (\ref{FRW-action-1}) will stay quadratic in the momentum operator $P_i$, independently of the coordinate frame one choses to work with. 
By using eq.~(\ref{operator-A-2}) instead of eq.~(\ref{operator-A}) back in the action  (\ref{FRW-action-1}) and defining $p = a(\tau) \rho$ we obtain:
\be
S =   \frac{1}{2} \int_{\beta a^{-2}p^2<1} \!\!\!\!\!\!   d \tau  \, d^3 p \, a^2 \left\{ \left|  \partial_{\tau}  \phi \right|^2   - \frac{ p^2 |\phi |^2}{(1 - a^{-2} \beta p^2)^2}   \right\}. \label{FRW-action-3}
\ee
Observe that the only difference between this form of the action and the one provided in Eq.~(\ref{FRW-action-2}) is the term involving time derivatives.\footnote{In particular, this prescription would restore the equivalence (up to a boundary term) of the action for the scalar perturbations expressed in terms of the Mukhanov variable and the curvature perturbations for non-zero $\beta$. This follows straightforwardly from the fact that the time derivative term in the Lagrangian for arbitrary $\beta$ is the same as it is in the case where $\beta = 0$.} Notice additionally, that here $p^{-1}$ denotes a comoving wavelength whereas $\rho^{-1} = a(\tau) p^{-1}$ corresponds to a physical wavelength. 
As we shall see in Section \ref{exact solutions}, this reformulation simplifies greatly the treatment of a scalar field on a de Sitter background which respects the GUP, as it allows to solve exactly for the mode functions in such a way that continuously deforms known results  in terms of $\beta$. But before we turn to this, we shall show that this prescription is in fact equivalent to working with a representation of the GUP formulated directly in position space, which we derive presently.

\subsection{Position space representation}

Strictly speaking our approach still corresponds to a choice of ordering in $\beta$. This in the sense that, as with (\ref{operator-A2}), we could have written  $\sum_{j} [\partial_{y^i}, y^j] \partial_{y^j}$ instead of $\partial_{y^i}$ in eq.~(\ref{FRW-phys-coord}), leading to extra terms in (\ref{FRW-action-3}) if we choose to proceed with the identification $ P_i = - i \partial_{y^i} $ and take $\beta \neq 0$. To clarify this and further justify our proposal, here we show that  our previous prescription is equivalent to a comoving space representation formulation of the generalized uncertainty principle, and therefore making unnecessary any reference to physical coordinates $y^i$.
Indeed, instead of (\ref{X-P-rho}), we wish to search for an alternative representation where the $X^i$ are diagonal. We begin this endeavor by parameterizing our canonical pair as
\bea
\hat X^i  \to  x^i, \qquad {\rm and} \qquad
\hat P_i  \to  -i\Omega(\beta\nabla^2)\nabla_i \label{alt}.
\eea
We start by working in a flat background (we generalize to FRW backgrounds further on), where $\nabla_i$ and $\nabla^2$ commute. In addition, we assume that $\Omega$ admits a covergent power series expansion 
\be
\label{omega}
\Omega(\beta\nabla^2) = \sum_{n=0}^{\infty} c_n(\beta\nabla^2)^n.
\ee  
We note from (\ref{alt}) that at the end of our calculations, we should expect to recover the standard representation of the momentum operator in the limit of momenta far less than the scale set by $1/\sqrt{\beta}$.
It remains then to reproduce (\ref{mod1}) with our ans\"atze (\ref{alt}). We find when acting on a scalar function $\phi$:
\be
\label{mcc}
[\hat X^i, \hat P_j]\phi = i\Omega(\beta\nabla^2)\delta^i_j\phi + i[\Omega(\beta\nabla^2),x^i]\nabla_j\phi.
\ee
Using the result $[\nabla^{2n},x^i]\nabla_j\phi = 2n\nabla^i\nabla_j \nabla^{2(n-1)}\phi$,
which is easily proved by induction after calculating for $n=1$, we find that 
\be
\label{om}
[\Omega(\beta\nabla^2),x^i]\nabla_j\phi = 2\beta\Omega'(\beta\nabla^2)\nabla^i\nabla_j\phi,
\ee
where $\Omega'$ denotes the derivative of the function $\Omega$ with respect to its argument. We then see that (\ref{mcc}) becomes 
\be
\label{mcc2}
[\hat X^i, \hat P_j] = i\Omega(\beta\nabla^2)\delta^i_j + i2\beta\Omega'(\beta\nabla^2)\nabla^i\nabla_j,
\ee
which we now have to equate to (\ref{mod1}) with $g = 2\beta$ and $f = 2\beta P^2/(\sqrt{1 + 4\beta P^2} - 1) = (\sqrt{1 + 4\beta P^2} + 1)/2$. Given that $\beta P^2 = -\Omega^2(\beta\nabla^2)\beta\nabla^2$, solving for $f$ implies
\be
\Omega(x) = \frac{1}{1 + x},
\ee
or equivalently:
\be
\label{final}
\hat P_j = \frac{-i\nabla_j}{1 + \beta\nabla^2}.
\ee
We see then that the remaining condition to satisfy (\ref{mod1}), $2\beta\Omega'(\beta\nabla^2)\nabla^i\nabla_j = 2\beta\hat P^i\hat P_j$ is also satisfied. Hence the prescription
\be
\label{pres}
-i\nabla_i \to \frac{-i\nabla_i}{1 + \beta\nabla^2},
\ee
is such that representing our canonical momentum thus furnishes a representation of the generalized uncertainty relations (\ref{mod1}). We note that in order for the operator $\hat P^i$ to be well defined, we must restrict ourselves in function space to functions composed of wave vectors such that $\beta k^2 < 1$ (this is equivalent to the requirement that (\ref{omega}) has a convergent power series expansion). That is, in Fourier space, we have a cut-off momentum $k < \beta^{-1/2}$, just as before. Hence we see that these two representations exist over the same function space. As we shall see shortly, there are considerable advantages in working with this representation which we persevere with. The definition of the operator (\ref{final}) should be clear-- it is defined through its power series (\ref{omega}) where the fact that all momenta are below the cutoff $\beta^{-1/2}$ ensures that this operator is well defined. In addition, we notice that the prescription (\ref{pres})
  introduces a spatial fuzziness around the scale $x \sim \sqrt\beta$, as roughly speaking this is when the higher derivatives become significant and introduce non-localities with the spatial extent $\sqrt{\beta}$.  

We now observe that this representation suggests an alternative way to specify the action for a scalar field on an FRW background, which is to stick to the comoving coordinate system $x^i$ and simply to use the prescription (\ref{pres}) directly in (\ref{FRW-com-coord}). In a spatially flat FRW spacetime, one has $[\nabla_i, \nabla_j] = 0$, which, just as in Minkowski spacetimes, eliminates any ambiguity in ordering $\nabla_{i}$. The resulting action is then given by
\be
S = - \frac 12 \int d\tau d^3 x \, a^2 (\tau) \left[  (\partial_{\tau} \phi)^2 - \sum_{i=1}^{3} \Big(\frac{\partial_i}{1 + \beta\nabla^2}\phi \Big)^2  \right], \label{FRW-com-coord-2}
\ee
where $\nabla^{2} = a^{-2} \sum_{i} \partial_{i}^2$. One can now rewrite this action in momentum space and obtain again the action of eq.~(\ref{FRW-action-3}). We see now that the requirement that our representation (\ref{pres}) be well defined translates into that we restrict ourselves to momenta such that $\sqrt{\beta} < a p^{-1}$, i.e. we only consider modes with a physical wavelength greater than the minimum length scale $\sqrt \beta$. Moreover, it is possible to verify either by using eq.~(\ref{FRW-action-3}) or eq.~(\ref{FRW-com-coord-2}) that there are no ambiguities now in choosing different gauges when considering cosmological perturbation theory as explored in~\cite{Ashoorioon:2004vm}.

\section{Exact Solutions} \label{exact solutions}

In the previous section we proposed an alternative prescription to~\cite{kempf1} to derive the dynamics of a scalar field theory on a de Sitter background. In this section we deduce the exact solutions to the mode equations resulting from~(\ref{FRW-action-3}) and briefly study their quantization.

The equation of motion deduced from~(\ref{FRW-action-3}) is given by:
\be
\phi'' - \frac{2}{\tau} \phi' +  \frac{ p^2 }{(1 - a^{-2} \beta p^2)^2} \phi = 0. \label{mot-1}
\ee
One can solve this equation analytically by setting $a^2 =  (H \tau)^{-2}$, where $H$ is the Hubble parameter taken to be a constant. 
The two independent solutions are
\be
\phi_{\pm}(\tau, p) =  \sqrt{1 - \beta H^2 p^2 \tau^2} \left[  1+ p \tau \left(  \beta H^2 p \tau \mp i \gamma \right) \right] e^{\pm i \frac{\gamma}{ \sqrt{\beta} H} {\rm arctanh} \left( \sqrt{\beta} H p \tau \right)  }, \label{sol-phi-1} 
\ee
where $\gamma = \sqrt{1 - \beta H^2}$. When the physical wavelength $\rho^{-1} = a p^{-1} \gg \sqrt{\beta}$ the solutions correspond to the usual modes of a free scalar field in a dS spacetime. Additionally, when $\rho^{-1} \gg H^{-1} > \sqrt{\beta}$, these modes become constant as expected.

We now would like to quantize the scalar field satisfying Eq.~(\ref{mot-1}). Towards this end we again impose the canonical commutation relation 
\bea
[\phi(\tau,p), \pi(\tau,p')] = i \delta^{3}(p - p'),
\eea 
where now $\pi(\tau,p) = a^2 \phi'(\tau,- p)$ is the canonical momentum deduced from Eq.~(\ref{FRW-action-3}). This can be done by writing the general solution to  Eq.~(\ref{mot-1}) in terms of the usual creation and annihilation operators
\be
\phi(\tau, \vec p) = \tilde \phi (\tau, \vec p) a^{\dag}(\vec p)  +  \tilde \phi^{*}(\tau, - \vec p) a(- \vec p) , \label{creation-op}
\ee
where $\tilde \phi (\tau, \vec p)$ is a linear combination of the two modes expressed in Eq.~(\ref{sol-phi-1}). One finds (up to an overall phase)
\bea
\label{alph}
\tilde \phi (\tau, \vec p) &=& \frac{H}{\sqrt{2 p^3}} u(\beta) \left[ \cosh \alpha~ \phi_{+}  +  e^{i\delta} \sinh \alpha ~\phi_{-} \right] ,   \\
\label{delt}
u(\beta) &=& \frac{1}{\sqrt{\gamma (1 + 3 \beta H^2)}}. 
\eea
Here $\alpha$ and $\delta$ are angles parametrizing the initial conditions of the perturbations.

One immediate feature to note, is that our solutions (\ref{sol-phi-1}) imply a stringy cut-off on the scale of inflation, namely
\be
\beta^{-1/2} > H.
\ee
One can verify that if $\beta^{-1/2} < H$, the solutions consist of only growing and decaying modes, which cannot be normalized and hence quantized. This is to be expected when we consider the stringy origins of the GUP, in that when the spacetime curvature approaches the string scale, effective field theory patently breaks down, signalled in our formalism by the non-normalizability of the mode functions at momenta above the string scale. 

In addition, we note that different values of $\alpha$ in (\ref{alph}) correspond to different choices of vacua. A simple choice would consist of $\alpha=0$. With this choice one recovers the usual modes of the Bunch-Davies vacuum for dS spacetimes when $ \sqrt{\beta} p a^{-1} \ll 1$. Typically, particle production bounds at the end of inflation and stability considerations have been used to rule out all values for $\alpha$ except $\alpha = 0$ \cite{kevin, al1, al2} when we consider inflation in the usual case. We will see how the presence of a cutoff on momentum space alters these considerations in the next section. 

In further analyzing the properties of these solutions one recovers many of the conclusions drawn in ref.~\cite{kempf2} and \cite{guptp} (for example, the divergence of the contribution of each mode to the energy density as $p \tau H \rightarrow \beta^{-1/2}$ addressed at length in the previous references), which is again to be expected as above anything else, this is signalling a breakdown of effective field theory at the string scale. Moreover, now that we have a relatively tractable set of mode functions to work with, we can explicitly calculate many quantities of phenomenological interest, not least of which is the two point correlation function, and compare them with observations to see what, if any signatures of the string scale parameterized by $\beta$, might be imprinted in the cosmic microwave background. 

\section{Comparison with observations}

We now concern ourselves with the observational consequences of inflation in the presence of the minimum length scale $\sqrt{\beta}$.
We start by noticing that the background field equations for a scalar field $\phi$ with potential $V(\phi)$ are unaffected by the presence of the length scale $\sqrt{\beta}$:
\be
\ddot \phi  + 3 H  \dot \phi +  \frac{\partial V}{\partial \phi} = 0.  \label{background-scalar}
\ee
This allows us to study the scalar fluctuations produced during inflation in the usual way, that is, by assuming that the inflaton field is slowly rolling and that $H$ remains almost unchanged during inflation. We can then compute the two point correlation function of a scalar field. Let us first consider the case where $\alpha=0$. We begin with (\ref{sol-phi-1}) and (\ref{creation-op}) and compute the two point correlation function for a scalar field on an exactly de Sitter background to be
\bea
\langle  \phi(p) \phi(p') \rangle = (2 \pi)^3 \delta^{3}(p + p') \frac{H^2}{2 p^3}  \frac{1}{\gamma (1 + 3 \beta H^2)},
\eea
where we have evaluated the amplitude at super horizon scales ($| p\tau | \ll 1$). From this result one can directly evaluate the two point correlation function for the curvature perturbation $\zeta$ on a slowly rolling background (recall that this quantity is constant on super horizon scales). The relation between $\zeta$ and $\phi$ is given by\footnote{See ref.~\cite{Ashoorioon:2004vm} for a discussion on how to justify the use of GUP on the curvature fluctuations $\zeta$.}
\be
\phi = - \frac{\dot \phi_0}{H} \zeta,
\ee
where $\phi_0$ is the vacuum expectation value of the inflaton field satisfying Eq.~(\ref{background-scalar}). Therefore, one has
\bea
\langle  \zeta(p) \zeta(p') \rangle = (2 \pi)^3 \delta^{3}(p + p') \frac{H_{*}^4}{2 p^3 \dot \phi_*^2}  \frac{1}{\gamma_* (1 + 3 \beta H_{*}^2)}. \label{two-point}
\eea
Where $H_*$ and $\gamma_*$ means that these quantities are to be evaluated when the mode $p$ crosses the horizon (i.e. when $ap^{-1} = H^{-1}$). We note that the spectral index $n_s$ of (\ref{two-point}) becomes
\bea
n_s -1 &=& p \frac{d}{d p} \ln \left(  \frac{H_{*}^4}{\dot \phi_*^2}  \frac{1}{\gamma_* (1 + 3 \beta H_{*}^2)} \right) \nonumber\\
&=&  2 (\eta - 3 \epsilon) -  p \frac{d}{d p} \ln \left( \gamma_* (1 + 3 \beta H_{*}^2) \right) \nonumber\\
&=& 2 (\eta - 3 \epsilon) - H_*^2 \beta \left[ \frac{1}{1 - \beta H_*^2} -  \frac{6}{1 + 3 \beta H_*^2}  \right] \epsilon, \label{spectral-ind}
\eea
where $\epsilon$ and $\eta$ are the usual slow roll parameters. In deducing the last expression, we used $p d/dp \simeq H_{*}^{-1} d/dt^{*}$. Note the order $\beta H^2 = H^2/m_s^2$ of the  corrections to the usual results in the above. Proceeding, we use (\ref{spectral-ind}) to rewrite Eq.~(\ref{two-point}) in terms of $p$. We do so by expressing this result in terms of the usual power spectrum for scalar perturbations $\mathcal{P}_{\mathcal{R}} (p)$:
\be
\label{pspect}
\mathcal{P}_{\mathcal{R}} (p) = \frac{1}{2 \epsilon} \left( \frac{H_*^{2}}{2 \pi}  \right) \frac{1}{\gamma_* (1 + 3 \beta H_{*}^2)} \left( \frac{p}{a H_*}  \right)^{n_s - 1}.
\ee
WMAP currently constrains $\mathcal{P}_{\mathcal{R}} (p)$ as
\be
\mathcal{P}_{\mathcal{R}} (p) \simeq 2.95 \times 10^{-9} A, \label{obs-power}
\ee
where $A = 0.6 - 1$ depending on the model. On the other hand one can also compute the two point correlation function for tensor perturbations. Here one finds 
\be
\mathcal{P}_{T} (p) =  8 \left( \frac{H_*^{2}}{2 \pi}  \right) \frac{1}{\gamma_* (1 + 3 \beta H_{*}^2)} \left( \frac{p}{a H_*}  \right)^{n_T - 1},
\ee
where 
\bea
n_T -1 =  - 2 \epsilon - H_*^2 \beta \left[ \frac{1}{1 - \beta H_*^2} -  \frac{6}{1 + 3 \beta H_*^2}  \right] \epsilon. 
\eea
From this result, the ratio of tensor to scalar perturbations is found to be the usual one: 
\be
r = 16 \epsilon.
\ee
We note here in passing that the usual consistency conditions for inflation coming from the  overdetermined nature of trying to relate three experimentally determined quantities ($r$, $n_s$ and $n_T$) to the two parameters $\eta$ and $\epsilon$, is modified\footnote{We thank Laura Covi for pointing this out to us.} by the presence of the additional parameter $\beta$ (although this effect is unlikely to be observable-- see our discussion further on).

Current constraints coming from WMAP are $r<0.22$ \cite{wmap5} which sets $\epsilon < 0.014$. Together with (\ref{obs-power}) provides the following bound on the scale of inflation $H$ taking into account $\beta$ (recall that we are working in units where $M_{Pl} = 1$)
\be
 \frac{H_*}{\sqrt{\gamma_* (1 + 3 \beta H_{*}^2)}}  < 2.3\sqrt A \times 10^{-5}.
\ee
This expression can be used to constrain the parameter space $(H,\beta)$. Observe that the possible observation of gravitational waves would severely  constrain $\beta$. In particular, a positive detection of the B-mode in CMB polarization, and therefore an indirect evidence of gravitational waves from inflation, once  foregrounds due to gravitational lensing from local sources have been properly treated, requires $\epsilon > 10^{-5}$ corresponding to (c.f.  \cite{Bartolo:2004if})
\be
\frac{H_*}{\sqrt{\gamma_* (1 + 3 \beta H_{*}^2)}}  > 1.3 \times 10^{-6}.
\ee

We note that a non-zero choice for $\alpha$ and $\delta$ in (\ref{alph}) would result in the two point correlation function
\eq{nonza}{\langle  \phi(p) \phi(p') \rangle = (2 \pi)^3 \delta^{3}(p + p') \frac{H^2}{2 p^3}  \frac{1}{\gamma (1 + 3 \beta H^2)}| \cosh \alpha - e^{i\delta} \sinh \alpha|^2,}
which simply rescales the two point correlation function. From (\ref{alph}), one can immediately infer that at the end of inflation, any non-zero value of $\alpha$ would result in a non-zero number occupation number of particles (with respect to the $\alpha = 0$ vacuum) given by:
\eq{modoc}{n_k = \sinh^2 \alpha.}
Since this expression is independent of $k$, we see that at the end of inflation in the usual case, any non-zero value of $\alpha$ would result in particle production at all wavelengths, which would singularly backreact on the geometry \cite{kevin}. However with the GUP, the finite cutoff implies the following energy density of scalar quanta at the end of inflation for non-zero values of $\alpha$: 
\eq{densend}{\rho = \frac{ \sinh^2 \alpha}{\beta^2} = m_s^4 \sinh^2 \alpha.}
Requiring that this quantity be small with respect to the energy density stored in inflation imposes the bound
\eq{infb}{ m_s^4 \sinh^2 \alpha \ll 3M^2_{Pl}H^2 \to \sinh \alpha \ll H M_{Pl}/m^2_s,}
which for GUT scale inflation implies $\sinh \alpha \ll 10^{-4}M^2_{Pl}/m_s^2$. Hence for a low enough string scale, it is straightforward to see how the bound pertaining to a non-trivial alpha parameter coming from particle overproduction is very slightly relaxed. However, the usual arguments concerning the instability of these non-trivial vacua still applies (see for instance \cite{kevin}\cite{brun}), hence we do not consider these vacua any further.

Having derived various corrections to quantities of interest that are observed in the CMB, it remains to put numbers on the magnitude of the effects. First we note from (\ref{obs-power}), that (\ref{pspect}) implies that inflation happened at, or around the GUT scale:
\eq{infscale}{H_* \simeq \epsilon^{1/2} 10^{15} \mathrm{GeV} ,}
which implies an energy scale $V^{1/4} \simeq \epsilon^{1/4}10^{16}$GeV. Given that the modifications arising from the GUP predict corrections of the order $\beta H^2 = H^2/m_s^2$, we see that in general, these corrections are of the order (for $\epsilon \sim 10^{-2}$):
\be
 M^2_{Pl}/m^2_s \times 10^{-8}.
 \ee
Given current experimental sensitivities, in order for these corrections to be detectable, we would require the corrections to be of the order of a percent (which also still ensures that our effective field theory treatment remains consistent), and thus would require a string scale of $m_s = 10^{15}$GeV, or an order of magnitude below the GUT scale. Even for reasonable assumptions for the string coupling such that we posit our universe to exist in a weakly coupled corner of moduli space, such a low string scale is exceedingly unlikely, and such corrections are likely to remain unobservable. However, a one order of magnitude improvement in CMB data will make us sensitive to the effects of GUT scale strings, as in this case the corrections would fall within experimental sensitivity. 

In concluding this section, we offer a comment on how our work relates to the wider literature on the so called `trans-Planckian problem' (see ~\cite{tp}-\cite{tp23} for a sampling). Clearly, the model we work with is such that the possibility of following the evolution of modes in super Planckian regimes is obviously precluded in the sense that such modes simply do not exist. The GUP implies a string scale (at weak coupling, a potentially sub-Planckian) cutoff on the allowed field modes, and provided they are created in the vacuum state corresponding to $\alpha = 0$ as the universe expands, leave an imprint that is unfortunately beyond the scope (though suggestively only by an order of a magnitude for GUT scale strings) of present experimental sensitivity. We now offer our concluding thoughts.

\section{Conclusions}

We discovered in this report that in working with the representation of the GUP given by (\ref{alt}), or alternatively (\ref{final}), it is possible to obtain exact solutions to the mode equations for a scalar field on a de Sitter background. We note that, consistent with the breakdown of effective field theory close to the string scale, such modes do not admit normalizable solutions if the scale of inflation is at or greater than the string scale $H > m_s$. This sensibly implies a string scale cutoff on the scale of inflation. Furthermore, we utilized our exact solutions to the mode equations to compute the two point correlation function of a massless scalar field on a de Sitter background, and computed corrections to the power spectrum and the tilt of the scalar and tensor spectra. We found that the corrections are far too small to be observed by current experimental sensitivities, although an order of magnitude improvement, would make observations sensitive to imprints of 
 the stringy minimal length at the GUT scale.

\section{Acknowledgements}
We would like to thank Robert Brandenberger and Laura Covi for comments on the draft, and many useful discussions, for which we also thank Jinn-Ouk Gong. We also thank Achim Kempf for discussions during early stages of this work. G.A.P. is supported by The Netherlands Organization for Scientific Research (N.W.O.) under the VICI and VIDI programmes, and by the German Science Foundation (DFG) under the Collaborative Research Centre (SFB) 676. S.P. wishes to thank Brian Greene and members of the ISCAP at Columbia University for hospitality during the preparation of this manuscript, and the DESY Hamburg theory group for hospitality during which this work was initiated. S.P. is supported at the Humboldt University in part by funds from project B5 of the SFB 647 (Raum Zeit Materie) grant, and is grateful to Alan Rendall at Albert Einstein Institute and Jan Plefka at the Humboldt University for this.

\end{document}